\documentclass[prl,twocolumn,preprintnumbers,amsmath,amssymb]{revtex4}
\usepackage{graphicx}
\usepackage{bm}

\begin{document}


\title{Voltage plateaus on $V(I)$ curves of long quasi-one-dimensional superconducting wires (without microwave irradiation)}


\author{V.\,I.~Kuznetsov}
\email{kvi@iptm.ru}
\author{A.\,A.~Firsov}
\affiliation{Institute of Microelectronics Technology and High
Purity Materials, Russian Academy of Sciences, Chernogolovka,
Moscow Region 142432, Russia}

\date{\today}

\begin{abstract}
Almost constant voltage plateaus on the $V(I)$ curves of long
quasi-one-dimensional superconducting aluminum wires placed in
magnetic field at temperatures $T$ slightly below the critical
superconducting temperature $T_{c}$ were found which were
unexpected for the sample geometry and parameters of the
experiment. The plateaus are assumed to be subharmonics of a
superconducting gap and arise due to the multiple Andreev
reflection and strong quasiparticle overheating in the wire
nonequilibrium  region. The plateaus are evidence of coexistence
of superconductivity and dissipation in such wires. The results
presented in the paper could not be described by existing
theories.
\end{abstract}

\maketitle
\section{INTRODUCTION}

Switching from the superconducting (S) to normal (N) state and
vice-versa and electron transport in a nonequilibrium resistive
state in an external current controlled superconducting wire of
small cross-section is a topical object for investigation. Such a
wire can be used as a single-photon detector \cite{gol'tsman} or a
key element in superconducting quantum interference device (SQUID)
\cite{barone}, phase-slip flux qubit \cite{mooij, astafiev}, and
highly efficient rectifier of ac voltage \cite{dubonosjetplett,
kuznetsovprb}.

It is known that SN(NS) switching in a superconducting
quasi-one-dimensional wire (with transverse dimensions smaller
than the doubled Ginzburg-Landau temperature-dependent coherence
length $2\xi(T)$) current-carrying occurs due to thermally
activated and quantum fluctuations of the superconducting  order
parameter leading to the appearance of the phase slip centers
(PSC) \cite{ivlev, zaikin}. In the PSC core of the $2\xi(T)$ size
the modulus of the order parameter periodically tends to zero
while its time-averaged value is non-zero. A thermally activated
phase slip center gives rise to a large number of nonequilibrium
quasiparticles which are distributed along the doubled length of
quasiparticle diffusion $2\lambda_{Q}(T)$ (external PSC size)
\cite{ivlev, schmidt}. A wire containing a phase slip center is in
the nonequilibrium state in which a superconducting and a
dissipative quasiparticle currents coexist.

The presence of quasiparticle current in a long wire with several
phase slip centers is supported by virtually linear sections on
the curve of current dependence of voltage $V(I)$ with a
differential resistance $dV/dI$ multiple to the resistance of one
phase slip center. The average direct superconducting current
(excess superconducting current $I_{exp}$) in the wire is
determined by the point where prolonged linear sections and the
current axis intersect \cite{schmidt}.

Constant voltage plateaus (integers with $n=1$ and the Shapiro
fractional steps) which appear at $V=mhf/2ne$ ($e$ is the
elementary charge, $n$ and $m$ are integers, $h$ is the Planck's
constant) on the $V(I)$ curves of superconducting structures with
a weak link \cite{shapiro, likharev} and quasi-one-dimensional
wires \cite{kuznetsovjetp} under microwave irradiation at the
frequency $f$ are evidence of alternating superconducting current
in the region of the phase slip center.

\begin{table*}
\caption{$L_{w}$ - total length of the current wire, $L_{n}$ -
length of the central part of the wire (or narrowing), $w_{w}$ -
width of the wide part of the wire, $w_{n}$ - width of the central
part of the wire (or narrowing), $d$ - film thickness, $R_{N}$ -
wire resistance in the N state, $l$ - mean free path of electrons
for wires}
\centering %
\begin{tabular}{cccccccccc}
\hline wire & $L_{w}$ ($\mu$m) & $L_{n}$ ($\mu$m) & $w_{w}$
($\mu$m) & $w_{n}$ ($\mu$m) & $d$ (nm) & $T_{c}$ (K) & $R_{N}$
($\Omega$)
& $\xi(0)$ ($\mu$m) & $l$ (nm)\\
\hline
WA &   70 & 8 & 0.46 & 0.46 & 32 & 1.407 & 17 & 0.14 & 16 \\
WB &   70 & 6.6 & 0.5 & 0.28 & 20 & 1.454 & 69 & 0.1 & 9 \\
WC &   70 & 0.9 & 1.6 & 0.28 & 29 & 1.355 & 4.4 & 0.12 & 13 \\
\hline
\end{tabular}
\end{table*}

A nonequilibrium electron transport in SNS junctions at
$V<2\Delta/e$ ($\Delta$ is the equilibrium value of  the
superconducting order parameter  (energy gap) in reservoirs)
undergoes modification due to the multiple Andreev reflection
(MAR) \cite{octavio, bardas, bezuglyi}. In a quasi-one-dimensional
superconducting wire, a phase slip center play the role of weak
link because the time-averaged modulus of the order parameter
(effective superconducting gap) in the core of the phase slip
center is smaller in value than the order modulus outside the core
of the phase slip center. During MAR, quasiparticles between two
superconducting reservoirs are reflected from time to time from
the reservoirs obtaining an energy of $neV>2\Delta$ which suffices
to leave the junction. Multiple crossing of the central junction
region by quasiparticles enhances nonequilibrium processes to
cause a strong quasiparticle diffusion and strong quasiparticle
(or Joule) overheating in the central region of the SNS junction
or the phase slip center. The MAR and quasiparticle overheating
\cite{octavio,bardas, bezuglyi, taboryski} lead to the appearance
of current singularities (subharmonics of the superconducting gap)
on $V(I)$ curves at $V=2\Delta/ne$.

MAR and gap subharmonics are described theoretically in the case
of ballistic junctions with mean free path for quasiparticles $l$
larger than the junction normal-region length $L$ \cite{octavio},
as well as in the case of diffusive SNS junctions ($l<<L$)
\cite{bardas, bezuglyi, taboryski}. The most pronounced
singularities can be expected in a short SNS junction
\cite{bardas}. The singularities are poorly visible on the $V(I)$
curves but are fairly distinct on the dependences of $dV/dI$ and
$dI/dV$ on $V$. The known theories expect that in
\cite{chennature} do not predict these singularities as voltage
plateaus on $V(I)$ curves.

Recently the authors of \cite{chennature} reported the observation
of almost constant-voltage plateau on $V(I)$ curves of short ($L$
length in the range $\xi(T)<<L<<\lambda_{Q}(T)$)
quasi-one-dimensional superconducting current-carrying wires
placed in magnetic field $B$ at $T$ much lower than $T_{c}$ . The
plateaus were due to MAR and quasiparticle overheating in the
phase slip center formed in the wire. The authors
\cite{chennature} claim "the universal character" of the states
with voltage plateaus because the plateau was observed for all
samples at $V_{0}\approx (0.43\pm 0.05)\Delta(0)/e$ (where
$\Delta(0)=1.764kT_{c}$ is the equilibrium superconducting energy
gap at $T=0$). The plateaus are evidence that a superconducting
and a dissipative quasiparticle currents coexist in the wire.

SN(NS) switching in short wires are measured in \cite{chennature}.
Though, as far as we know there's been no studies of SN(NS)
switching in the case of direct current flowing through a rather
long (total length $L_{w} \approx 6-10\lambda_{Q}(T)$)
quasi-one-dimensional superconducting wire, while $V$ has been
taken from a center short (with length $L_{n}$ satisfying a
condition $\xi(T)<L_{n}<\lambda_{Q}(T)$) part of the wire.

In this work we present the measured $V(I)$ curves of three
relatively long wires of different geometries (the sketches of the
wire central part are given in the insets of Figs. \ref{f1},
\ref{f3}, and \ref{f4}) placed in the field $B$ perpendicular to
the substrate surface at $T$ slightly below $T_{c}$. The wires
were prepared by thermal deposition of a thin aluminum film onto a
silicon substrate by the lift-off technique of electron-beam
lithography. Thickness and surface of the wires was checked by an
atomic force microscope. The wire parameters are given in the
table.
\section{SAMPLES AND EXPERIMENTAL PROCEDURE}
A direct current $I_{dc}$ through $I_{1}$  and $I_{2}$  wires (the
insets of Figs. \ref{f1}, \ref{f3}, and \ref{f4}) was applied to
70 $\mu$m long current wires $w_{w}$ wide which had either no
narrowing (wire WA) or a short narrowing $w_{n}$  wide in the wire
center (wires WB and WC). Voltage $V$ was measured in the central
part of the wire between the $V_{1}$ and $V_{2}$  wires (the
insets of Figs. \ref{f1}, \ref{f3}, and \ref{f4}). The potential
wires were 70 $\mu$m long with the width equal to the width of the
wire central part. The large length of current and potential wires
allows the decrease of the effect of 15 $\mu$m wide
superconducting leads (which continue the wires) on the electron
transport in the wire center.

According to the theories known to these authors, theory in
\cite{chennature} included, gap subharmonics cannot be expected to
reveal themselves as voltage plateaus on the $V(I)$ curves for the
wires at measurement parameters used in this experiment. However,
a number of plateaus of non-universal character were found, which
suggests the coexistence of excess superconducting and dissipative
quasiparticle currents in a relatively long wire.

The value of $l$ was determined from a refined theoretical
expression \cite{gershenson} $\rho l=5.1\times 10^{-16}$ $\Omega$
$m^{2}$ where $\rho$ is the wire resistivity. Because $l<<\xi_{0}$
for the wires studied (where $\xi_{0}=1.6$ $\mu$m is the length of
superconducting coherence of pure aluminum at $T=0$\,K), the
Ginsburg-Landau length at $T$ slightly below $T_{c}$  was
calculated by expression $\xi(T)=\xi(0)(1-T/T_{c})^{-1/2}$ (here
$\xi(0)=0.85(\xi_{0}l)^{1/2}$) \cite{schmidt}. In the temperature
range of the experiment the wire was quasi-one-dimensional and
$\lambda_{Q}(T)=7-12$\,$\mu$m.

\section{RESULTS AND DISCUSSION}
\begin{figure}
\includegraphics[width=1\linewidth]{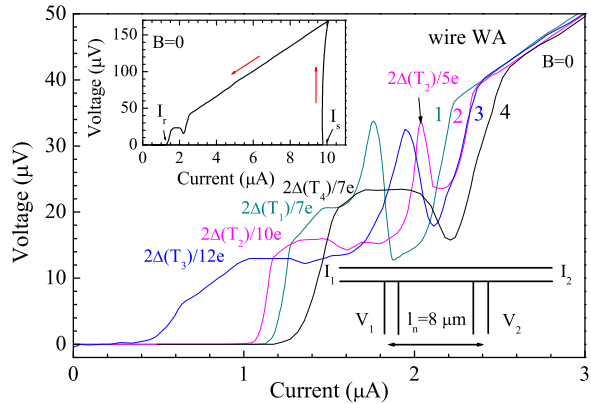}
\caption{\label{f1} (Color online) $V(I)$ curves for the WA wire,
measured with decreasing current in $B=0$ at $T$ :  $T_{1}=1.347$
- curve 1, $T_{2}=1.344$ - 2, $T_{3}=1.341$ - 3, $T_{4}=1.340$\,K
- 4. The critical temperature is $T_{c}=1.407$\,K. The
corresponding voltages $V_{pl}=2\Delta(T)/ne$ ($n=7$, 10, 12, and
7 for curves 1-4, respectively) are given near the plateau. The
arrow indicates the maximum at $V=2\Delta(T_{2})/5e$. The insert
at the top: the $V(I)$ curve, recorded with increasing and
decreasing current in $B=0$ at $T=1.34$\,K; the insert at the
bottom: the sketch of the central part of the WA wire (not to
scale).}
\end{figure}

\begin{figure}
\includegraphics[width=1\linewidth]{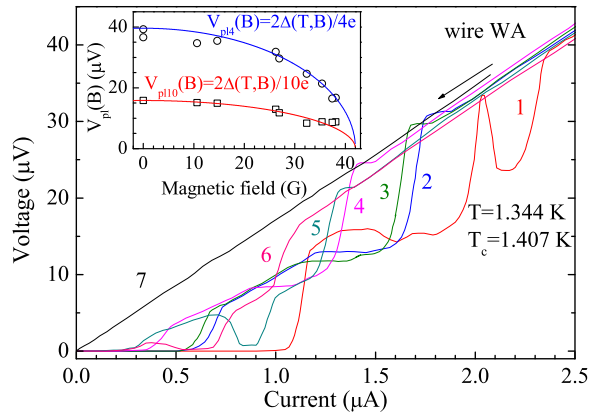}
\caption{\label{f2} (Color online) $V(I)$  data for the WA wire,
measured at $T=1.344$\,K in the magnetic field $B$: 0 - curve 1,
26.2 - 2, 26.9 - 3, 32.3 - 4, 35.4 - 5, 37.4 - 6,
49.4\,G\,-\,curve 7. The critical temperature is $T_{c}=1.407$\,K.
The insert: $V_{pl}(B)$ functions. The experimental data (circles
and squares) are approximated by theoretical curves (lines)
$2\Delta(B,T)/ne$ (where $n$ is equal to 4 and 10 for circles and
squares, respectively).}
\end{figure}

\begin{figure}
\includegraphics[width=1\linewidth]{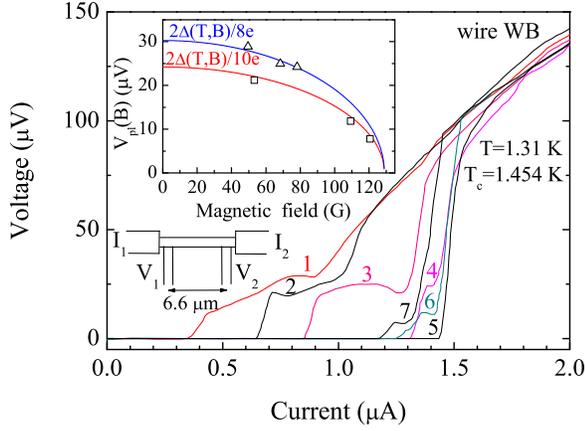}
\caption{\label{f3} (Color online) $V(I)$ curves for the WB wire,
recorded at $T=1.31$\,K in the fields $B$: 49.6 - curve 1, 53.2 -
2, 68.4 - 3, 78.1 - 4, 82.6 - 5, 109.3 - 6, 120.5\,G - curve 7.
The critical temperature is $T_{c}=1.454$\,K. The plateaus
correspond to the voltages $2\Delta(B,T)/8e$ (for curves 1, 3, and
4) and $2\Delta(B,T)/10e$ (for curves 2, 6, and 7). The left
insert: Triangles and squares - measured $V_{pl}(B)$ functions.
Lines - the theoretical curves $2\Delta(B,T)/ne$ with $n=8$ and 10
approximate the measured data. The insert at the bottom: the
sketch of the WB wire (not to scale).}
\end{figure}

\begin{figure}
\includegraphics[width=1\linewidth]{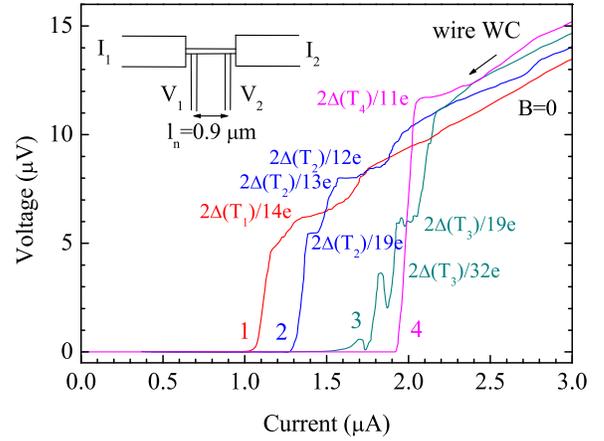}
\caption{\label{f4} (Color online) $V(I)$ curves for the WC wire
in $B=0$ at $T$ : $T_{1}=1.334$ - curve 1, $T_{2}=1.327$ - 2,
$T_{3}=1.320$ - 3, $T_{4}=1.312$\,K - 4. The critical temperature
is $T_{c}=1.355$\,K. The corresponding voltages
$V_{pl}(T)=2\Delta(T)/n$ are given near the plateau. The insert:
the WC wire sketch.}
\end{figure}

It was found that the $V(I)$ curves in a general case exhibit a
hysteresis depending on the direction of current sweep. The $V(I)$
curve of the WA wire measured at $B=0$ and $T=1.34$\,K displays
two superconducting critical currents: a switching current $I_{s}$
at which $V$ appears in the wire central part as $I$ increases and
a retrapping superconducting current $I_{r}$ at which $V$
disappears with decreasing current $I$ (the upper inset of Fig.
\ref{f1}). It was also found that $I_{s}$ is equal to the
Ginsburg-Landau deparing current $I_{GL}$ and exceeds $I_{r}$
which has a smaller value due to a quasiparticle (or Joule) wire
overheating. The hysteresis in the $V(I)$ curves disappears in
strong fields. In the zero field (the upper inset of Fig.
\ref{f1}) and weak fields the SN switching occurs abruptly as
current increases due to strong overheating of the wire. The
object of this study was an extended NS switching. Figures
\ref{f1}-\ref{f4} (except the upper inset of Fig. \ref{f1}) shows
the branches of $V(I)$ curves corresponding to NS switching.

Figure \ref{f1} presents the $V(I)$ curves for the WA wire
measured at $B=0$ and $T$ slightly below $T_{c}=1.407$\,K. At the
beginning of the NS switching, the $V(I)$ curves exhibit a
non-monotonic behavior as maxima (curves 1-3). This behavior is
caused by the competition of overheating and cooling of the
nonequilibrium region in the wire center. The maximum corresponds
to an excess quasiparticle overheating. Earlier, $V(I)$ curves
with overheating effects were also observed in aluminum
microbridges \cite{klapwijk}.

Moreover, $V(I)$ curves (Fig. \ref{f1} and the inset of Fig.
\ref{f1}) display steps of almost constant voltage (voltage
plateaus). The voltage $V_{pl}$ at which plateaus occur depends on
$T$. As the current $I$ decreases the plateau usually transforms
into a virtually linear section which ends by an abrupt drop to a
state close to the superconducting state. We believe that the
excess superconducting current and linear sections on the $V(I)$
curves of the studied wires as well as on the $V(I)$ curves of the
wires reported in \cite{kuznetsovjetp} evidently suggest the
presence of the phase slip center in the wire central part. The
$V(I)$ curves were found independent of the thermocycling: curves
1 and 2 were measured in one cycle, curves 3 and 4 in another
cycle (Fig. \ref{f1}). Figure \ref{f1} shows that the $V(I)$
curves (curves 3 and 4) measured at virtually equal temperatures
reveal certain lack of coincidence when current and potential
wires interchange their places (curve 4).

We supposed that the plateaus arise due to MAR and quasiparticle
overheating in the phase slip center or a short SNS junction
formed in the wire center during SN(NS) switching. In our case of
a long wire, phase slip center and a self-forming short SNS
junction differ in that the time-averaged gap in the SNS junction
center is smaller than the gap in the core of the phase slip
center because of strong overheating of the SNS junction. It was
expected that a plateau would appear at voltages
$V_{pl}=2\Delta(T,B)/ne$, where $\Delta(T,B)$ is the almost
equilibrium superconducting gap depending on $T$ and $B$ taken in
the region outside the PSC core or the SNS junction center. This
region in wires with narrowing corresponds to the place where
narrow parts of current and potential wires intersect. Note that
current singularities in SNS junctions are usually determined by
$\Delta(T,B)$, taken in the region of wide superconducting leads
\cite{octavio, bardas, bezuglyi}. The gap for the
quasi-one-dimensional Al superconducting wire was found from the
expression $\Delta(T,B)=\Delta(T)(1-(B/B_{mth}(T))^{2})^{1/2}$,
where $\Delta(T)=1.74\Delta(0)(1-T/T_{c})^{1/2}$ is the
temperature-dependent gap in the zero field at $T$ slightly below
$T_{c}$ and $B_{mth}(T)$ is the calculated temperature-dependent
maximum field at which the gap becomes zero \cite{schmidt}.

In Fig. \ref{f1} the voltage plateaus, considered to be
subharmonics of the gap, are observed at $V_{pl}=2\Delta(T)/ne$
(where $T$ are temperatures corresponding to curves 1-4, $n$ is
defined as an integer close to fitting parameter and equal to 7,
10, 12, and 7 for curves 1-4, respectively). Thus, the plateaus on
the curves can correspond to different values of $n$. In addition,
splitting of the plateau into two plateaus with close $V_{pl}$
voltages is seen on the curves 2, and 3. A splitting in
subharmonic particularities (not in a form of a plateau) of the
gap corresponding to $n=2$, and 4 was found earlier in diffusive
SNS structures \cite{taboryski}.

Figure \ref{f2} presents the $V(I)$ curves for the wire WA
measured in different fields at $T=1.344$\,K. One of the curves
exhibits up to two plateaus which are the gap subharmonics. We
believe that the plateaus seen in the beginning of the NS
switching are supposed to arise due to the appearance of a short
self-forming SNS junction in the wire center whereas other
plateaus arising at lower currents are caused by phase slip
centers. As current $I$ decreases the plateau caused by the phase
slip center becomes almost linear segment. As the field increases,
the peak on the $V(I)$ curves (curve 1) related to the beginning
of the NS switching disappears because the field reduces the gap
in the region outside the wire center to cause a more efficient
cooling of overheated quasiparticles \cite{vodolazov}.
The character of the curves again becomes non-monotonic in high
fields in the region close to the S state (curves 5 and 6). The
maxima on curves 5 and 6 are related to the reappearance of the
phase slip center with a current decrease, which had earlier
disappeared in higher fields. At $B=37.4$\,G (curve 6) close to
the maximum field $B_{m}(T=1.344\,K)=42$\,G, both plateaus
disappear because of a strong decreasing of the gap. At $37.4$\,G
the gap is not equal to zero yet therefore the phase slip center
remains in the wire center (as is evidenced by the linear section
of curve 6 and an excess superconducting current in the region
close to the S state (curve 6)). Thus, the formation of a phase
slip center or a SNS junction in the wire center at certain
current (voltage) values is an essential condition for plateau
observation.

The inset of Fig. \ref{f2} suggests that the plateau (Fig.
\ref{f2}) at voltages $V_{pl}(B)$ depending on field can be
regarded as the gap subharmonics with $n=4$ and $n=10$. It is seen
that the plotted $V_{pl}(B)$ functions (circles and squares) taken
from the experimental $V(I)$ curves ($T=1.344$\,K) of the WA wire
can be fitted by the theoretical curves
$V_{pl4}(B)=2\Delta(T,B)/4e$ and $V_{pl10}(B)=2\Delta(T,B)/10e$,
where $\Delta(T,B)=\Delta(T)(1-(B/B_{m})^{2})^{1/2}$ and
$B_{m}=42$\,G is the only fitting parameter for the maximum
magnetic field close to the calculated value of
$B_{mth}(T=1.344\,K)=38$\,G.

Figure \ref{f3} shows the $V(I)$ curves for the WB wire (with
narrowing) measured in different fields at $T=1.31$\,K. As the
field increases the retrapping superconducting current $I_{r}$
increases (curves 1-5, a similar anomalous behavior of $I_{r}$ was
found in the "magnetic-field-induced superconductivity" and
"anti-proximity effects" \cite{chenprl,tian}); at further
increasing of the field $I_{r}$ demonstrates a usual decreasing
(curves 6, and 7). One of the curves exhibits a short horizontal
segment of almost constant voltage or one slightly tilted segment.
We regard both the segments as plateaus. One curve exhibits one
plateau. The plateaus depend on field and are observed (curves
1-4, 6-7) up to the fields close to the calculated maximum
magnetic field in the wire narrowing $B_{mnth}(T=1.31\,K)=129$\,G.
The plateau is not observed on curve 5 recorded in the field close
to the maximum field in the wide part of the wire
$B_{mwth}=73$\,G.

The inset of Fig. \ref{f3} shows that plateaus (Fig. \ref{f3}) at
$V_{pl}(B)$ can be treated as the gap subharmonics with $n=8$, and
$n=10$. It is seen that the plotted $V_{pl}(B)$ (triangles and
squares) data derived from the measured $V(I)$ curves
($T=1.31$\,K) for the WB wire fit the theoretical curves
$V_{pl8}(B)=2\Delta(T,B)/8e$ and $V_{pl10}(B)=2\Delta(T,B)/10e$,
where $\Delta(T,B)=\Delta(T)(1-(B/B_{mnth}(T))^{2})^{1/2}$ at
$T=1.31$\,K.

Figure \ref{f4} exhibits the $V(I)$ curves for the WC wire
recorded in the zero field at temperatures $T$ slightly below
$T_{c}=1.355$\,K. Short almost horizontal segments along with
slightly tilted segments of approximately constant voltage
(plateaus) are seen on the $V(I)$ curves. One curve can display
not less than three plateaus (curve 2). We found the $V_{pl}$  is
a function of the temperature. We believe that the plateaus (Fig.
\ref{f4}) can be treated as the gap subharmonics $2\Delta(T)/n$
with $n=11$, 12, 13, 14, 19, and 32. The specific features of the
WC wire are a short narrowing of the $L_{n}<2\xi(T)$ length in the
studied range and a large width (1.6$\mu$\,m) of the major wide
part of the wire. The wire was quasi-one-dimensional at
$T>1.32$\,K. At this geometry, a nonequilibrium region in the
narrowing is more appropriate to regard as Josephson (a short SNS)
junction rather than a phase slip center \cite{likharev}.

The formation of a SNS junction in the wire center instead of a
phase slip center is supported by the fact that plateaus
disappeared at mutual exchange of current and potential wires.
Apart from plateaus, at high voltages $V$, when a state of the
wire narrowing is approaches to the a normal state and the rest of
the wire remains superconducting, we can observe a number of
subharmonic singularities (not in a form of a plateau) on $V(I)$
curves (the right part of Fig. \ref{f4}) and $dI/dV(V)$ curves
(not shown here). For instance $dI/dV$ as a function of $V$ with
the parameters of the $V(I)$ curve (curve 4 in Fig. \ref{f4})
shows a successive array of the singularities corresponding to
$n=3-11$, where the singularity with $n=11$ is attributed to the
plateau on the $V(I)$ curve. Singularities with $n >11$ are
invisible because they relate to the vertical segment of the
$V(I)$ curve.

A question could be asked: why could we not see a full successive
array of subharmonics (starting with $n=1$) and why are
subharmonics that are revealed as plateaus so scarce or why could
we see a plateau corresponding to a large subharmonic number
($n=32$).

We answer as follows. Because the $V(I)$ curves of our wires were
recorded in a bias direct current mode, that is why instead of
subharmonic singularities that are observed at low $V$ in the
region of negative values for $dV/dI$ in a bias voltage mode, we
see almost vertical segments of the $V(I)$ curves. At higher $V$
values in the resistive region close to the normal state, a
successive subharmonic array might also not be seen as SNS
junction appearance is improbable due to an overheating of the
wire center and of the adjacent parts of the wire close to the
center. That is true with the exception of the WC wire center
(Fig. \ref{f4}) with wide adjacent parts that remain
superconducting when measured in the zero field.

We found experimentally that plateaus are registered in a
restricted voltage interval relating to a nonequilibrium part on
$V(I)$ curves, when a steady SNS junction or a phase slip center
is appeared in the middle of the wire, as is verified by a linear
segment of the $V(I)$ curve. Only subharmonics with given numbers
$n$ (visible as plateaus at voltages $V_{pl}=2\Delta(T,B)/ne$) get
into this interval. The upper boundary of this interval
$V<2\Delta(T,B)/e$ is defined by multiplication of the wire
resistance in normal state by the value of the current at which
the wire resistance begins sharply drop with decreasing current. A
considerable decrease of the $V$ due to wire overheating leads to
a decrease in maximum possible value of $V_{pl}(B)$ and
consequently to the increase of a maximum number $n$ of the
visible subharmonics.

To observe multiple Andreev reflection with large numbers $n$ in
diffusive SNS contacts, it is necessary that $l_{in}$
\cite{bardas} inelastic scattering length (in our case, $l_{in}$
corresponds to twice the length of quasiparticle diffusion
$2\lambda_{Q}(T) \approx 14-24$\,$\mu$m) is much longer than the
length of the normal region. This requirement, written as
$l_{in}=2\lambda_{Q}(T)>>2\xi(T)$, is fulfilled here.

We believe subharmonic features, including plateaus, arise due to
MAR and quasiparticle overheating in a short SNS contact or in a
phase slip center. To observe the plateau, a stronger overheating
is necessary, which can be realized in aluminum and zinc wires due
to the long electron-phonon relaxation times. So these relaxation
times for aluminum and zinc equal to
$\tau_{E}=1.3\times10^{-8}$\,s and $\tau_{E}=9.3\times10^{-8}$\,s,
respectively, and exceed by a factor of two the relaxation time
for tin $\tau_{E}=9.3\times10^{-11}$\,s \cite{stuivinga}.

The coupling of the alternating Josephson current with the
external resonator, randomly formed in the measuring electrical
circuit, or the internal resonator in the wire, cannot cause a
plateau in the observed voltage range of 5-40 $\mu$\,$V$ (Figs.
\ref{f1}-\ref{f4}). Since the interval of Josephson frequencies
corresponding to this voltage range does not intersect with the
expected frequency intervals of the external and internal
resonators. In addition, it could be expected that the voltage, at
which a plateau occurs, caused by resonators will not clearly
depend on $T$ and $B$, whereas in our case the voltage at which a
certain plateau is observed depends on $T$ and $B$.

Below the comparison is given of the results obtained in this work
with those reported in \cite{chennature}.

1. In the zero field, our $V(I)$ curves had a hysteresis depending
on the direction of the current sweep, while the switching current
$I_{s}$  was equal to $I_{GL}$  and the retrapping current $I_{r}$
could have values 8-15 times smaller than $I_{GL}$ . Whereas in
\cite{chennature} $V(I)$  curves were not hysteresis, while
$I_{s}$ and the retrapping current $I_{r}$  coincided and were
30-50 times smaller than $I_{GL}$. The very low $I_{s}/I_{GL}$
value means that the electron transport in wires \cite{chennature}
is strongly influenced by internal fluctuations and external
noises, leading to premature wire switching with increasing
external current from the S state to the N state and the absence
of a $V(I)$ curve hysteresis. The very low value of $I_{r}/I_{GL}$
indicates a strong overheating of the current-carrying wire
\cite{chennature}, leading to a later (at low currents) wire
switching with a decrease in the current from the N state to the S
state. So, fluctuations, external noises and overheating are
stronger in wires \cite{chennature} than in our wires.

2. The voltage plateau on the $V(I)$ curves of our wires was
observed at $T$ slightly below $T_{c}$. The curves of
\cite{chennature} exhibit the plateau at $T$ much lower than
$T_{c}$.

3. The plateaus were stable with time. In \cite{chennature} the
plateaus were reported bistable. The bistability, in our opinion,
arises because of strong internal fluctuations, external noises
and overheating on the wires of \cite{chennature}.

4. We could observe as many as two plateaus on one $V(I)$ curve
(Figs. \ref{f2}, and \ref{f4}) caused by MAR both in the SNS
junction and phase slip center. The experiment and theory in
\cite{chennature} display only one plateau due to MAR in the phase
slip center.

5. One of the important differences between the results of this
work and those in \cite{chennature} is the following. The states
with plateau observed in this work are not universal because the
voltage $V_{pl}(T,B)=2\Delta(T,B)/ne$ at which the plateau arises
depends on $T$, $B$, and the gap subharmonic gap number $n$.
Plateaus were observed which correspond to large $n$ (up to
$n=32$). The authors of \cite{chennature} state that states with
plateau are of universal character. The theory proposed in
\cite{chennature} predicts that the plateau can be observed at
voltages $V_{0}^{th}=2\Delta_{eff}/e \approx 0.34\Delta(0)/e$
independent of $T$ , $B$, and $n$, where $\Delta_{eff}$ is an
average nonequilibrium suppressed gap in the wire center. The
excess of the experimental ratio $V_{0}/(\Delta(0)/e)$ over the
theoretical ratio $V_{0}^{th}/(\Delta(0)/e)=0.34$ reaches $35\%$
for the zinc wire (Sample A) and $44\%$ for the aluminum wire
(Sample D) \cite{chennature}. Thus, there is a difference between
the theory \cite{chennature} and the measurements
\cite{chennature}.

We propose an alternative calculation of the voltage at which
there is a plateau in \cite{chennature}, using the expression
$V_{pl}(T,B)=2\Delta(T,B)/ne$ for non-universal plateaus from our
work. In the general case, the value of the temperature-dependent
gap in the zero field should be obtained from the
Bardeen-Cooper-Schrieffer theory and not from the expression
$\Delta(T)$, valid near $T_{c}$ \cite{schmidt}. It is easy to
check that for the wires of the Sample A and Sample D
\cite{chennature}, the values of $V_{pl}(T,B)$ calculated within
our model at $T=0.45$\,K coincide with the measured values
\cite{chennature} with an accuracy of $1\%$ if  $n=4$. Thus, the
plateaus \cite{chennature} can be understood as the gap
subharmonics with $n=4$.

Since the plateau \cite{chennature} disappears in weak fields
$B>20$\,G and near Tc, the "universality" of the plateau (i.e.,
the independence of the voltage value at which the plateau is
observed from $T$ and $B$) cannot be experimentally verified for a
wide range of temperatures and fields. This "universality" has
been experimentally proven \cite{chennature} with an accuracy of
$\approx 12\%$ for a narrow field interval $B<20$\,G (where the
expected change in the gap in our model is $\approx 1\%$) and the
temperature interval $0<T<0.64T_{c}$ (where the expected change in
the gap in our model is $\approx 12\%$).

6. The field suppression of the gap in wide wires far from the
wire center did not affect the plateau. The plateau was observed
in a wide range of fields. The plateau on the $V(I)$ curves of the
WB wire (with narrowing) was observed when the gap in the main
wide part of the wire was set to zero and disappeared when the gap
in the wire narrow part was zero. In the WA wire (without
narrowing) the plateau disappeared when the gap was set to zero in
the region outside the SNS junction or in the PSC core. In
\cite{chennature} the range of fields at which the plateau exists
was very narrow because of a strong effect of wide superconducting
leads. The plateau disappeared \cite{chennature} when the gap in
the superconducting lead slightly decreased due to a weak field or
when $T$ approached $T_{c}$.

7. The $V(I)$ curves of the WA wire (without narrowing) displayed
the plateau in the absence of effects similar to those described
in \cite{chenprl, tian}. Moreover, the $V(I)$ curves of the WB
wire (with narrowing) exhibited the plateau both in the presence
of the effects described in \cite{chenprl, tian} (curves 1-4, Fig.
\ref{f3}) and without such effects (curves 5-7, Fig. \ref{f3}).
This disagrees with the statement \cite{chennature} that the
appearance of the state with a voltage plateau is strongly
connected with these effects \cite{chenprl, tian}.

\section{CONCLUSION}
To sum up, we observed a new effect which proves the coexistence
of superconductivity and dissipation in the nonequilibrium state
of relatively long quasi-one-dimensional current-currying aluminum
wires. The effect observed radically differs from the effect
reported in \cite{chennature}. In this work non-universal plateaus
of almost constant voltage on the $V(I)$ curves for the wires used
which were recorded at a decreasing current in magnetic field at
$T$ slightly below $T_{c}$  were found. These plateaus were quite
unexpected for these wires at the external parameters of the
experiment. One of the curves could exhibit as many as four
plateaus. Commonly, subharmonic features of the superconducting
gap caused by multiple Andreev refelection and quasiparticle
overheating in the nonequilibrium region display themselves as
current singularities on the $V(I)$ curves.

We found that these subharmonic features of the gap can appear as
plateaus at voltages $V_{pl}(T,B)=2\Delta(T,B)/ne$, with $n$
having certain integer values depending on $T$, $B$, and $V$. We
believe that \emph{a very strong quasiparticle overheating} in the
PSC core or SNS junction, caused by \emph{a long time of
electron-phonon relaxation in aluminum wires}, is necessary for
the plateau to appear. The observation of the plateau depends on
the wire geometry. For example, the plateau exists up to strong
fields close to the maximum fields $H_{m} \propto w_{w}^{-1}$ for
wires of constant width $w_{w}$ and $H_{mn} \propto w_{n}^{-1}$
for wires with narrowing $w_{n}$, respectively. Splitting of the
plateau (Fig. \ref{f1}) and non-monotone character of $V(I)$
curves were observed upon current variation caused switching of
the wire between different states as a result of competition
between quasiparticle overheating and cooling in the wire center.
In the region close to the N state subharmonics are seen in Fig.
\ref{f4} as weak current singularities (not plateaus).

\section{ACKNOWLEDGMENTS}
The work was supported by the Presidium of Russian Academy of
Sciences (research program "Challenging Problems of Low
Temperature Physics"). We thank V.\,Tulin, D.\,Vodolazov,
A.\,Melnikov, M.\,Skvortsov for fruitful discussions and
O.\,Trofimov for technical assistance.

\end{document}